\documentclass{new_tlp}

\usepackage{xcolor}
\usepackage{xifthen}
\usepackage{tikz-qtree}
\usepackage{etex}
\usepackage{amsmath}
\usepackage{amssymb}
\usepackage{proof}
\usepackage[hyphens,spaces]{url}

\usepackage{hyperref}
\usepackage{caption}
\usepackage{comment}
\usepackage{subcaption}
\usepackage{fancybox}
\usepackage{stmaryrd}
\usepackage[T1]{fontenc}
\usepackage[utf8]{inputenc}

\usepackage{bussproofs} 
\usepackage{empheq}

\usepackage[disable]{todonotes}

\usepackage{amstext}

\usepackage{local-macros}
\usepackage{references}

\usepackage{cleveref}
\crefformat{footnote}{#2\footnotemark[#1]#3}

\newtheorem{example}{Example}[section]
\newtheorem{definition}{Definition}[section]

\newtheorem{theorem}{Theorem}[section]
\newtheorem{lemma}{Lemma}[section]

\newcommand*{\mybox}[1]{\ovalbox{#1}}

\newcounter{reviewer}
\newcounter{point}[reviewer]
\renewcommand{\thepoint}{P\,\thereviewer.\arabic{point}}

\newcommand{\shortreply}[2][]{\medskip \noindent \sffamily\textbf{Reply}:\ #2
 \ifthenelse{\equal{#1}{}}{}{ \hfill \footnotesize (#1)}%
 \medskip}

\title[We Cannot Eliminate Cuts, But We Can Explore Them]{The New Normal: We Cannot Eliminate Cuts in Coinductive Calculi, But We Can Explore Them}

\author[E. Komendantskaya and D. Rozplokhas  and H. Basold]
{Ekaterina Komendantskaya\\
Heriot-Watt University, Edinburgh, Scotland, UK\\
\email{ek19@hw.ac.uk}
\and
Dmitry Rozplokhas\\
Jet Brains Research, St Petersburgh, Russia\\
\email{rozplokhas@gmail.com}
\and
Henning Basold\\
Leiden University, the Netherlands\\
\email{henning@basold.eu}
}

\begin{document}
\maketitle
\begin{abstract}
  In sequent calculi, cut elimination is a property that guarantees that any provable formula can be
  proven analytically.
  For example, Gentzen's classical and intuitionistic calculi \LK{} and \LJ{} enjoy cut elimination.
  The property is less studied in coinductive extensions of sequent calculi.
  In this paper, we use coinductive Horn clause theories to show that cut is not eliminable in a
  coinductive extension of \LJ{}, a system we call \CLJ{}.
  We derive two further practical results from this study.
  We show that CoLP by Gupta et al. gives rise to cut-free proofs in \CLJ{} with fixpoint terms, and
  we formulate and implement a novel method of coinductive theory exploration that provides several
  heuristics for discovery of cut formulae in \CLJ{}.
\end{abstract}


\section{Introduction}
\label{sec:intro}\label{sec:ex}

Cut elimination is one of the central properties of interest for sequent calculi~\cite{Gentzen69},
and more generally, proof theory.
Informally, whenever we want to prove a formula $\varphi$ relative to a given theory $\Gamma$, we
can use cut to first prove another formula $\psi$, and then show that $\psi$ implies $\varphi$:

\vspace*{-1em}

{\footnotesize
\begin{empheq}[box=\fbox]{gather*}
  \def\ScoreOverhang{1pt}
      \def\defaultHypSeparation{\hskip .15in}
      \def\labelSpacing{2pt}
      \def\ScoreOverhang{1pt}
      \def\labelSpacing{2pt}
   \AxiomC{$\Gamma \vdash \psi$}
      \AxiomC{$\Gamma, \psi \vdash \varphi$}
      \RightLabel{\Cut}
      \BinaryInfC{$\Gamma \vdash \varphi$}
      \bottomAlignProof
      \DisplayProof
\end{empheq}}

The \emph{cut elimination property} holds if every proof of a sequent that uses a cut,
can be transformed into a cut-free proof.
Cut elimination serves as a form of completeness result for the calculus:
cut-free proofs can be constructed analytically by simply following the structure of formulae,
eliminating any need to discover a cut formula.
For first-order logic, the  most famous example of a calculus with eliminable cut is Gentzen's system \LJ{}.

Recently, coinduction became a prominent proof method, and has been incorporated into a number of
proof systems, see e.g. \cite{BrotherstonS11} for cyclic proof systems or \cite{GuptaBMSM07} for
coinductive logic programming (CoLP).
Informally speaking, coinductive extensions of proof systems give finitary methods to prove
formulae that would otherwise require an infinite proof.
Usually, it comes in the shape of a fixpoint rule:

\vspace*{-1em}

{\footnotesize
\begin{empheq}[box=\fbox]{gather*}
  \def\ScoreOverhang{1pt}
      \def\defaultHypSeparation{\hskip .15in}
      \def\labelSpacing{2pt}
      \def\ScoreOverhang{1pt}
      \def\labelSpacing{2pt}
      \AxiomC{$\Gamma, \varphi \vdash \varphi$}
      \RightLabel{$\Cofix$}
      \UnaryInfC{$\Gamma \vdash \varphi$}
      \bottomAlignProof
      \DisplayProof
    \end{empheq}}
\vspace*{-0.5em}

The rule allows one to add the formula $\varphi$, that would otherwise cause infinite derivations,
directly to the set of assumptions, and thus close the proof coinductively in a finite number of
steps.
Sometimes $\varphi$ is called a \emph{coinduction hypothesis}.
The $\Cofix$ rule usually comes with certain guardedness or productivity conditions.
These vary from system to system, but always serve to guarantee soundness of the rule.

Only recently, the relation between these two principles, cut elimination and coinduction, attracted
special attention of the proof-theoretic community.
A series of papers~\cite{SNK20,KNTU20} showed that cut is not eliminable in a cyclic first-order
Separation logic.
In this paper, we show that this problem is more general:
Adding a coinduction rule to a first-order proof system destroys the property of cut elimination.
We show this for the Gentzen's intuitionistic sequent calculus \LJ{}, although a similar argument works  for Gentzen's \LK{}, and any sequent calculus for a logic with
implication and universal and existential quantification.
We call \LJ{} augmented with the cofix rule \emph{Coinductive \LJ{}}, or simply \CLJ{}, and show that
cut is not eliminable in \CLJ{}.

The system \CLJ{} is very similar to \emph{coinductive uniform proofs} (CUP)~\cite{BasoldKL19-2},
only that CUP does not feature a cut rule.
CUP is a coinductive extension of uniform proofs, a fragment of the Gentzen's sequent
calculus introduced to model the derivations obtained by first-order resolution
in Prolog~\cite{MNPS91,MN12}. 
As it turns out, CUP is sound with respect to the largest Herbrand models of logic programs~\cite{BasoldKL19-2}.

We apply our result in two ways.
Firstly, we show that derivations in CoLP~\cite{GuptaBMSM07} in fact correspond to
cut-free proofs in \CLJ{}.
This gives a proof-theoretic characterisation to the well-known results of incompleteness of CoLP.
Moreover, our characterisation of CoLP's loops by fixpoint terms may pave the way for future
embeddings of CoLP in richer theorem provers.

Secondly, seeing that we cannot hope to prove all theorems of interest analytically,
we propose to establish a stronger infrastructure for \emph{theory exploration} in coinductive first-order theories.
Similarly to the \emph{Boyer-Moore Waterfall Model}~\cite{Boyer79:ComputationalLogic}, the methodology consists of four steps:
(1) use a suitable coinductive sequent calculus (e.g. \CLJ{} without cut or CUP) to prove analytically as much as possible;
(2) use first-order resolution to explore the loops in derivations and suggest suitable coinductive lemmas;
(3) use the calculus to prove the discovered lemmas and discard those that cannot be proven;
(4) use the proven lemmas as cut formulae to complete previously failed proofs.

We present an implementation of this method, that comprises an implementation of CUP, several coinductive theory exploration methods from the literature, including CoLP and the method of
\citeN{FKS15}, as well as one novel theory exploration method.  The implementation is available on Github\footnote{\label{fn:impl}\url{https://github.com/CoUniform/theory-exploration}}.
These results are of interest to either logic programmers who need to reason about richer coinductive properties  than CoLP already handles, or the developers of other theorem provers that feature coinduction.

We can illustrate this paper's results by means of three examples.

\paragraph{Cut Non-Eliminability.}
Consider the following logic program $\Gamma_T$:
\begin{equation*}
  \kappa_{u} : \all{x} p(f(x))   \impl p(x) \, ,
\end{equation*}
and the goal formula $p(a)$ for some constant $a$.
We may attempt to prove $p(a)$ by means of an infinite tree that follows the rules of the system \LJ{}:

{\footnotesize{
\begin{prooftree}
  \AxiomC{$\vdots$}
  \RightLabel{\rulelabel{$\forall$-L}}
      \UnaryInfC{$\Gamma_T \vdash p(f(a))$}
      \AxiomC{$p(a) \equiv p(a)$}
      \RightLabel{\LJAxiom}
      \UnaryInfC{$\Gamma_T , p(a) \vdash p(a)$}
      \RightLabel{\rulelabel{$\rightarrow$-L}}
      \BinaryInfC{$\Gamma_T , p(f(a))  \rightarrow p(a)\ \vdash p(a)$}
      \RightLabel{\rulelabel{$\forall$-L}}
      \UnaryInfC{$\Gamma_T,\Gamma_T \vdash p(a)$}
      \RightLabel{\rulelabel{C-L}}
      \UnaryInfC{$\Gamma_T \vdash p(a)$}
    \end{prooftree}}}

In fact, $p(a)$ is not directly (analytically) provable in \LJ{}.
However, if we proved the lemma $\all{x} p (x)$, we could derive $p(a)$ as an instance.
Such a proof for $p(a)$ in our system \CLJ{} is shown in \Cref{fig:CLJ}.
Sequents in \CLJ{} have contexts that consist of three parts that are separated by ``$+$'':
the logic program $\Gamma_T$, a context with ordinary proof assumptions (see the application of the
rule {\footnotesize{$\LJImplLT$}}), and one which holds coinduction hypotheses (see the application
of the rule {\footnotesize{$\LJCoFixRule$}}).
This splitting of contexts allows us to ensure \emph{guardedness}, and therefore soundness of
coinductive proofs.
The proof proceeds by introducing $\all{x} p(x)$ through the cut rule into the proof of $p(a)$ in
the lower part of \Cref{fig:CLJ}.
We then proceed to prove $\all{x} p (x)$ by using the {\footnotesize\LJCoFixRule{}}-rule, and we
therefore call this formula a coinduction hypothesis.

\begin{figure}
    \footnotesize
\begin{prooftree}
      \AxiomC{$p(x) \equiv p(x)$}
      \RightLabel{\LJAxiom}
      \UnaryInfC{$\SequentCLJ(\{p(x)\})[\emptyset][\{\all{x} p(x)\}]{p(x)}$}
      \AxiomC{$p(f(x)) \equiv p(f(x))$}
      \RightLabel{\LJAxiom}
      \UnaryInfC{$\SequentCLJ(\emptyset)[\{p(f(x))\}][\emptyset]{p(f(x))}$}
      \RightLabel{\LJAllLG}
      \UnaryInfC{$\SequentCLJ(\emptyset)[\{\all{x} p(x)\}][\emptyset]{p(f(x))}$}
      \RightLabel{\LJImplLT}
      \BinaryInfC{$\SequentCLJ(\{p(f(x)) \impl p(x)\})[\emptyset][\{\all{x} p(x)\}]{p(x)}$}
      \RightLabel{\LJAllLT}
      \UnaryInfC{$\SequentCLJ[\emptyset][\{\all{x} p(x)\}]{p(x)}$}
      \RightLabel{\LJAllR}
      \UnaryInfC{$\SequentCLJ[\emptyset][\{\all{x} p(x)\}]{\all{x} p(x)}$}
      \RightLabel{\LJCoFixRule}
      \UnaryInfC{$\spadesuit$}
\end{prooftree}
\vspace*{1em}
\begin{prooftree}
      \AxiomC{$\spadesuit$}
      \UnaryInfC{$\SequentCLJ[\emptyset][\emptyset]{\all{x} p(x)}$}
      \AxiomC{$p(a) \equiv p(a)$}
      \RightLabel{\LJAxiom}
      \UnaryInfC{$\SequentCLJ[\{p(a)\}][\emptyset]{p(a)}$}
      \RightLabel{\LJAllLG}
      \UnaryInfC{$\SequentCLJ[\{\all{x} p(x)\}][\emptyset]{p(a)}$}
      \RightLabel{\LJCut}
      \BinaryInfC{$\SequentCLJ[\emptyset][\emptyset]{p(a)}$}
\end{prooftree}
  \caption{\emph{\footnotesize{A coinductive proof in \CLJ{} with cut.}}}
  \label{fig:CLJ}
\end{figure}

In \Cref{sec:cut}, we will use this example to prove cut non-eliminability in \CLJ{}.
That is, we will show that it is impossible to give a cut-free proof for
$\SequentCLJ[\emptyset][\emptyset]{p(a)}$.
It is worth noting that coinductive inference for $p(a)$ also cannot be accomplished in
CoLP~\cite{GuptaBMSM07}, and this logic program has been used to show incompleteness of CoLP.

\paragraph{Understanding the Proof-Theoretic Power of CoLP.}
Looking with proof-theoretic spectacles at CoLP, we notice that CoLP requires circular unifiers seen as
fixpoint terms to represent rational terms but does not require the cut rule.
For example, consider the logic program $P_{\stream0}$ that defines the  stream of zeros:
\begin{equation*}
  \kappa_{\stream 0} :
  \all{x }  \stream(x)  \impl \stream(\scons(0, x))
\end{equation*}

CoLP finds a loop in the resolution trace
$\stream(x) \stackrel{x / \scons(0, x)}{\rightarrow} \stream(x) \rightarrow \ldots $,
and generates a circular unifier $x = \scons(0, x)$ as a finitary representation of the stream.
The Prolog query $\stream(x)$ corresponds to the goal $\exist{t}\; \stream(t)$ in \CLJ{}.
In order to obtain a proof for $P_{\stream} + \emptyset + \emptyset \vdash \exist{t}\; \stream(t)$
in \CLJ{}, we will need to instantiate the existential variable $t$
with the term $s := \fix[x] \scons(0, x)$.
Note the use of a fixpoint at the term level as an alternative representation for circular unifiers.
We can then prove $P_{\stream} + \emptyset + \emptyset \vdash  \stream(s)$
by {\footnotesize{$\LJCoFixRule$}} with $\stream(s)$ as coinduction hypothesis.
More generally, all CoLP proofs yield cut-free proofs in \CLJ{}, as we will show in \Cref{sec:exist}.

\paragraph{Going Beyond State of the Art.}
The above results allow us to look at the picture more generally, and notice that proofs of some propositions in coinductive first-order Horn clause theories
in fact require proving coinduction lemmas that are formulated in a richer language.
Already in our simple example,
$\all{x} p (x)$ is a goal in hereditary Harrop logic, rather than Horn clause logic because
universal goals cannot be proven in Prolog.
One can find examples when \emph{higher-order} coinductive lemmas are needed to complete proofs
arising from logic programs.
Take, for example, the logic program $P_{\fromP}$ that defines
streams of successive natural numbers, e.g., $0, s(0), s(s(0)), \cdots$ :
\begin{equation*}
  \kappa_{\fromP} :
  \all{x \, y} \fromP(s(x), y) \impl \fromP(x, \scons(x, y))
\end{equation*}
To prove the goal $\exist{t} \fromP(0, t)$,
we have to find a finitary representation of the (infinite) term
$\scons(0, \scons(s(0), \dotsm)$.
This is not possible with circular unifiers, but rather with \emph{higher-order} fixpoint terms.
Moreover, we also have to generalise our goal, which leads to the coinduction lemma
$\all{x} \fromP (x, \fix[f] \lam{x} \scons(x, f(s \, x)))$.
From this lemma, we are able to obtain $\exist{t} \fromP(0, t)$ as a corollary.

In order to prove lemmas at this level of generality, one could use $\lambda$-Prolog~\cite{MN12}
that features both higher-order terms and hereditary Harrop clauses.
CUP~\cite{BasoldKL19-2} shows that a coinductive extension of $\lambda$-Prolog is sound relative
to the greatest Herbrand models.
However, CUP itself has no capacity to \emph{search} for lemmas that can serve as coinduction hypotheses,
it can only \emph{prove} one correct if it is already found.
In \Cref{sec:irreg}, we contribute several theory exploration techniques.
Coinductive theory exloration for the example $\Gamma_T$ from above has already been introduced in~\citeN{FKS15}.
Our implementation incorporates this method, the CoLP-style search for fixpoint terms, and one novel extension that
also searches for higher-order coinduction hypotheses, as required for the example $P_{\fromP}$.

\section{Background: Fixpoint Terms and Horn Clause theories}
\label{sec:bg}

We will only work with first-order Horn clause theories in this paper.
However, in presence of coinduction, even these theories may require formulae with higher-order fixpoint terms,
as we saw in the introduction.
This motivates the use of simply typed $\lambda$- and fixpoint-terms~\cite{Barendregt:LambdaCalcTypes,BasoldKL19-2}.
For Horn clause theory definitions, we
follow closely the notation used in Uniform proofs~\cite{MN12}.

We define the sets $\Types$ of \emph{types} and
$\PropT$ of \emph{proposition types}
by the following grammars, where $\baseT$ and $\propT$
are the \emph{base type} and \emph{base proposition type}.
\begin{equation*}
  \Types \ni \sigma, \tau \coloncolonequals
  \baseT
  \mid \sigma \to \tau
  \qquad \qquad
  \PropT \ni \rho \coloncolonequals
  \propT
  \mid \sigma \to \rho, \quad \sigma \in \Types
\end{equation*}

A \emph{term signature} $\TSig$ is a set of pairs $c : \tau$, where
$\tau \in \Types$, and a \emph{predicate signature} is a set
$\PSig$ of pairs $p : \rho$ with $\rho \in \PropT$.
The elements in $\TSig$ and $\PSig$ are called \emph{term symbols}
and \emph{predicate symbols}, respectively.
Given term and predicate signatures $\TSig$ and $\PSig$, we refer to
the pair $(\TSig, \PSig)$ as \emph{signature}.
Let $\Vars$ be a countable set of variables, the elements of which we denote
by $x, y, \dotsc$
We call a finite list $\Gamma$ of pairs $x : \tau$ of variables and types a
\emph{context}.
The set  $\FixTerms{\TSig}$ of \emph{(well-typed) terms} over $\TSig$
is the collection of all $M$ with $\typed{M}{\tau}$ for some context
$\Gamma$ and type $\tau \in \Types$, where $\typed{M}{\tau}$ is defined
 in \Cref{fig:terms}.
A term is called \emph{closed} if $\typed[]{M}{\tau}$, otherwise it is called
\emph{open}.
We say that $\varphi$ is a \emph{(well-formed) formula} in context $\Gamma$,
if $\validForm{\varphi}$ is inductively derivable from the rules
in \Cref{fig:terms}.

It is customary in logic programming to write the arguments to symbols as tuples like,
for example, in $f(t_{1}, t_{2})$.
Our definition uses juxtaposition instead for simplicity, that is, we would write this term as
$f \, t_{1} \, t_{2}$.
Throughout this paper, we will, however, often employ the logic programming style for the benefit
of the reader.

\begin{figure}[!]
  \footnotesize{
    \begin{spreadlines}{7pt}
      \begin{empheq}[box=\fbox]{gather*}
        \AxiomC{$c : \tau \in \TSig$}
        \UnaryInfC{$\typed{c}{\tau}$}
        \DisplayProof
        \quad
        \AxiomC{$x : \tau \in \Gamma$}
        \UnaryInfC{$\typed{x}{\tau}$}
        \DisplayProof
        \quad
        \AxiomC{$\typed{M}{\sigma \to \tau}$}
        \AxiomC{$\typed{N}{\sigma}$}
        \BinaryInfC{$\typed{M \> N}{\tau}$}
        \DisplayProof
        \\
        \AxiomC{$\typed[\Gamma, x : \sigma]{M}{\tau}$}
        \UnaryInfC{$\typed{\lam{x} M}{\sigma \to \tau}$}
        \DisplayProof
        \quad
        \AxiomC{$\typed[\Gamma, x : \tau]{M}{\tau}$}
        \UnaryInfC{$\typed{\fix[x] M}{\tau}$}
        \DisplayProof
      \end{empheq}
    \end{spreadlines}
    \begin{spreadlines}{7pt}
      \begin{empheq}[box=\fbox]{gather*}
        \def\defaultHypSeparation{\hskip .05in}
        \AxiomC{$(p : \tau_1 \to \dotsm \to \tau_n \to \propT) \in \PSig$}
        \AxiomC{$\typed{M_1}{\tau_1}$}
        \AxiomC{$\dotsm$}
        \AxiomC{$\typed{M_n}{\tau_n}$}
        \QuaternaryInfC{$\validForm{p \> M_1 \dotsm \> M_n}$}
        \bottomAlignProof
        \DisplayProof
        \\[7pt]
        \AxiomC{$\validForm{\varphi}$}
        \AxiomC{$\validForm{\psi}$}
        \AxiomC{$\Box \in \set{\conj, \disj, \impl}$}
        \TrinaryInfC{$\validForm{\varphi \mathbin{\Box} \psi}$}
        \bottomAlignProof
        \DisplayProof
        \quad
        \AxiomC{$\validForm[\Gamma, x:\tau]{\varphi}$}
        \UnaryInfC{$\validForm{\all{x:\tau} \varphi}$}
        \bottomAlignProof
        \DisplayProof
        \quad
        \AxiomC{$\validForm[\Gamma, x : \tau]{\varphi}$}
        \UnaryInfC{$\validForm{\exist{x : \tau} \varphi}$}
        \bottomAlignProof
        \DisplayProof
      \end{empheq}
    \end{spreadlines}
    \vspace*{-1em}
    \caption{\emph{\footnotesize{\textbf{Top:} Well-formed Terms. \textbf{Bottom:} Well-formed Formulae.}}}
    \label{fig:terms}}
\end{figure}

We will use a standard $\beta$- and $\fix$-reduction relation on terms, see~\cite{BasoldKL19-2}.
The equivalence closure of the reduction relation (convertibility) is denoted by $\conv$.

 The \emph{order} of a type $\tau \in \Types$ is given as usual
by $\ord(\iota) = 0$ and
$\ord(\sigma \to \tau) = \max\set{\ord(\sigma) + 1, \ord(\tau)}$.
If $\ord(\tau) \leq 1$, then the arity of $\tau$ is given by
$\ar(\iota) = 0$ and $\ar(\iota \to \tau) = \ar(\tau) + 1$.
A signature $\TSig$ is called \emph{first-order}, if for all $f : \tau \in \TSig$
we have $\ord(\tau) \leq 1$; similarly for $\PSig$.
We let the arity of $f$ then be $\ar(\tau)$ and denote it by $\ar(f)$.

The \emph{guarded base terms} over a first-order signature
$\TSig$ are given by the following rules.

{\footnotesize{\begin{gather*}
      \AxiomC{$x : \tau \in \Gamma$}
      \AxiomC{$\ord(\tau) \leq 1$}
      \BinaryInfC{$\guarded{x}{\tau}$}
      \DisplayProof
      \quad
      \AxiomC{$f : \tau \in \TSig$}
      \UnaryInfC{$\guarded{f}{\tau}$}
      \DisplayProof
      \quad
      \AxiomC{$\guarded{M}{\sigma \to \tau}$}
      \AxiomC{$\guarded{N}{\sigma}$}
      \BinaryInfC{$\guarded{M \> N}{\tau}$}
      \DisplayProof
      \\
      \def\defaultHypSeparation{\hskip .1in}
      \AxiomC{$f : \sigma \in \TSig$}
      \AxiomC{$\ord(\tau) \leq 1$}
      \AxiomC{$
        \guarded[\Gamma, x : \tau, y_1 : \iota, \dotsc, y_{\ar(\tau)} : \iota]{
          M_i}{\iota}$}
      \AxiomC{$1 \leq i \leq \ar(f)$}
      \QuaternaryInfC{$\guarded{\fix[x] \lam{\vec{y}} f \> \vec{M}}{\tau}$}
      \DisplayProof
    \end{gather*}}}
General \emph{guarded terms}
are generated
by the following grammar.
\begin{equation*}
  G \coloncolonequals
  M \; (\text{with } \guarded[]{M}{\tau} \text{ for some type } \tau)
  \mid c \in \TSig \mid x \in \Vars \mid G \; G \mid \lam{x} G
\end{equation*}
Finally, $M$ is a \emph{first-order} term over
$\TSig$ with $\typed{M}{\tau}$ if $\ord(\tau) \leq 1$ and the types of all
variables occurring in $\Gamma$ are of order $0$.

Note that an important aspect of guarded terms is that no free variable occurs
under a $\fix$-operator.
\emph{Guarded base terms} should be seen as specific fixpoint terms
that we will be able to unfold into potentially infinite trees.
\emph{Guarded terms} close guarded base terms under operations of the
simply typed $\lambda$-calculus. \citeN{BasoldKL19-2} provides examples and further discussion of guarded terms.
In what follows, we will use the following sets of well-typed terms:
the set  $\Terms{\TSig}$ of all \emph{simple terms}, i.e. terms that
do not involve $\fix$;
the  set $\GuardedFOTerms{\TSig}$  of guarded first-order terms;
the set $\FOSimpleTerms{\TSig}$ of simple first-order terms.

\begin{definition}[Atoms]
  \label{def:atoms}
  A formula $\varphi$ of the shape $p \; M_1 \dotsm \; M_n$ is an \emph{atom} and a
  \begin{itemize}
  \item \emph{first-order atom}, if $p$ and all the terms $M_i$ are first-order;
  \item \emph{guarded atom}, if all terms $M_i$ are guarded; and
  \item \emph{simple atom}, if all terms $M_i$ are simple.
  \end{itemize}
 The sets of first-order, guarded and simple atoms are denoted by
  $\foAt$, $\guardedAt$ and $\simpleAt$.
  We denote intersections of these sets by
  $\foGuardedAt = \foAt \cap \guardedAt$ and
  $\foSimpleAt = \foAt \cap \simpleAt$.
\end{definition}

\begin{definition}[D- and G-formulae, Logic Programs, Coinduction Hypothesis]
  \label{def:formula-classes}
  Let $D$ and $G$ be generated by the following grammar.
  \begin{alignat*}{2}
    D & \coloncolonequals
    \guardedAt \mid G \impl D \mid  D \conj D \mid \all{x : \tau} D  \\
    G  & \coloncolonequals
    \guardedAt  \mid G \conj G \mid  G \disj G
    \mid \exist{x : \tau} G \mid D \impl G \mid \all{x  : \tau} G
  \end{alignat*}

  A $D$-formula of the shape
  $\all{\vec{x}} A_1 \conj  \dotsm \conj  A_n \impl A_0 $
  is called \emph{$H$-formula} or \emph{Horn clause} if $A_k \in \foSimpleAt$.
  Finally, a \emph{logic program} (or \emph{program}) $P$ is a set of
  $H$-formulae.

  A formula $\varphi$ is a
  \emph{coinduction hypothesis}  if $\varphi$ simultaneously is a
  $D$- and a $G$-formula.
\end{definition}

$D$- and $G$-formulae are also known as \emph{definite clauses} and \emph{goal clauses} in the logic
programming literature.
The above syntax of $D$ and $G$-formulae  in fact presents an extension of Horn clause syntax to
hereditary Harrop formulae (that allow universal and implicative goals).
Coming back to our running example of $\Gamma_T$ formulated in \Cref{sec:ex}, we see  that
$\Gamma_T$ was given by a Horn clause.
However, the proof of a goal $p(a)$ required to prove $\all{x} p(x)$ first, which is a goal of
hereditary Harrop logic.

\section{Coinductive Sequent Calculus \CLJ{}; Proof of Cut Non-Elimination}\label{sec:cut}

We start with introducing \CLJ{}, a coinductive dialect of the Gentzen's intuitionistic sequent calculus \LJ{}~\cite{Gentzen69}.
The rules in \Cref{fig:rules-CLJ} follow the standard formulation of \LJ{}~\cite{SU06} (including notation $\Gamma, \psi$ for $\Gamma \cup \{\psi\}$), except for the following three differences.
Firstly, we restrict ourselves to logic programs for $\Gamma_T$, and we allow only $G$-formulae in $\Gamma_A$ and $\Gamma_C$.
As a result, we omit some \LJ{} rules for existential and disjunctive formulae on the left.
Secondly, we introduce the rule {\footnotesize{$\LJCoFixRule$}} in its standard formulation, see e.g.~\cite{BasoldKL19-2}.
Finally,  we ensure guardedness of coinduction in \CLJ{} by splitting the context into logic programs $\Gamma_T$, intermediate proof assumptions $\Gamma_A$, and
coinduction assumptions $\Gamma_C$. Applying the rule {\footnotesize{$\LJCoFixRule$}} is the only way of introducing a coinduction assumption in $\Gamma_C$.
But, to complete a proof that starts with {\footnotesize{$\LJCoFixRule$}}, we can never use formulae from $\Gamma_C$.
The only rule that allows us to shift the coinduction hypotheses from $\Gamma_C$ to $\Gamma_A$ and thus make them usable in proofs is the rule  {\footnotesize{$\LJImplLT$}}. Inuitively, this means we can only use a coinduction assumption after we ``resolved'' our current goal against some clause from $\Gamma_T$.

For this section only, it is sufficient to take a much smaller fragment of \CLJ{}, and restrict ourselves to only simple first-order atoms in $\Gamma_A$ and $\Gamma_C$. In later sections, it will be made clear how and why higher-order and fixpoint terms can be useful.

\LJ{} has four structural rules: weakening, exchange and contraction on the left, and weakening on the right.
We omit the latter, as we extend the {\footnotesize{$\LJAxiom$}} rule in a way that renders right weakening opaque.
To mimic \LJ{}, we need to add the remaining three structural rules {\footnotesize{\LJWeakLT{}}}, {\footnotesize{\LJExchLT{}}}
and {\footnotesize{\LJCtrLT{}}} for $\Gamma_T$:

{\footnotesize
  \begin{gather*}
    \AxiomC{$\SequentCLJ(\ThAssumsEnv){\varphi}$}
    \UnaryInfC{$\SequentCLJ(\ThAssumsEnv, \psi){\varphi}$}
    \bottomAlignProof
    \DisplayProof
    \qquad
    \AxiomC{$\SequentCLJ(\ThAssumsEnv, \phi, \psi, \Gamma_T'){\varphi}$}
    \UnaryInfC{$\SequentCLJ(\ThAssumsEnv, \psi, \phi, \Gamma_T'){\varphi}$}
    \bottomAlignProof
    \DisplayProof
    \qquad
    \AxiomC{$\SequentCLJ(\ThAssumsEnv, \psi, \psi){\varphi}$}
    \UnaryInfC{$\SequentCLJ(\ThAssumsEnv, \psi){\varphi}$}
    \bottomAlignProof
    \DisplayProof
  \end{gather*}}
and similarly for $\Gamma_A$.
We assume these 6 rules additionally to those in \Cref{fig:rules-CLJ}.

We do not state soundness of \CLJ{} here, as soundness of a very similar proof system CUP relative to the greatest Herbrand models of logic programs
was already proven in \cite{BasoldKL19-2}.
Here, our main goal is to prove cut non-elimnation in \CLJ{}. We use the example of \Cref{sec:ex} to show this.

\begin{figure}[bt]
  \footnotesize{
  \begin{spreadlines}{7pt}
    \begin{empheq}[box=\fbox]{gather*}
      \def\ScoreOverhang{1pt}
      \def\defaultHypSeparation{\hskip .15in}
      \def\labelSpacing{2pt}
      \def\ScoreOverhang{1pt}
      \def\labelSpacing{2pt}
      \AxiomC{$\varphi' \in \ThAssumsEnv \cup \GuardedAssumsEnv$}
      \AxiomC{$\varphi \conv \varphi'$}
      \RightLabel{\LJAxiom}
      \BinaryInfC{$\SequentCLJ{\varphi}$}
      \bottomAlignProof
      \DisplayProof
      \quad
      \AxiomC{$\SequentCLJ{\varphi_1}$}
      \AxiomC{$\SequentCLJ{\varphi_2}$}
      \RightLabel{\LJConjR}
      \BinaryInfC{$\SequentCLJ{\varphi_1 \conj \varphi_2}$}
      \bottomAlignProof
      \DisplayProof
      \\
      \AxiomC{$\SequentCLJ(\ThAssumsEnv, \psi_i){\varphi}$}
      \AxiomC{$i \in \{1, 2\}$}
      \RightLabel{\LJConjLT}
      \BinaryInfC{$\SequentCLJ(\ThAssumsEnv, \psi_1 \conj \psi_2){\varphi}$}
      \bottomAlignProof
      \DisplayProof
      \quad
      \AxiomC{$\SequentCLJ[\GuardedAssumsEnv, \psi_i]{\varphi}$}
      \AxiomC{$i \in \{1, 2\}$}
      \RightLabel{\LJConjLG}
      \BinaryInfC{$\SequentCLJ[\GuardedAssumsEnv, \psi_1 \conj \psi_2]{\varphi}$}
      \bottomAlignProof
      \DisplayProof
      \\
      \AxiomC{$\SequentCLJ{\varphi}$}
      \AxiomC{$x \not\in FV(\ThAssumsEnv \cup \GuardedAssumsEnv \cup \CoindAssumsEnv)$}
      \RightLabel{\LJAllR}
      \BinaryInfC{$\SequentCLJ{\all{x}\varphi}$}
      \bottomAlignProof
      \DisplayProof
      \quad
      \AxiomC{$\SequentCLJ{\varphi\subst{N/x}}$}
      \RightLabel{\LJExR}
      \UnaryInfC{$\SequentCLJ{\exist{x}\varphi}$}
      \bottomAlignProof
      \DisplayProof
      \\
        \AxiomC{$\SequentCLJ(\ThAssumsEnv, \psi\subst{N/x}){\varphi}$}
      \RightLabel{\LJAllLT}
      \UnaryInfC{$\SequentCLJ(\ThAssumsEnv, \all{x}\psi){\varphi}$}
      \bottomAlignProof
      \DisplayProof
      \quad
      \AxiomC{$\SequentCLJ[\GuardedAssumsEnv, \psi\subst{N/x}]{\varphi}$}
      \RightLabel{\LJAllLG}
      \UnaryInfC{$\SequentCLJ[\GuardedAssumsEnv, \all{x}\psi]{\varphi}$}
      \bottomAlignProof
      \DisplayProof
      \\
      \AxiomC{$\SequentCLJ[\GuardedAssumsEnv, \psi]{\varphi}$}
      \RightLabel{\LJImplR}
      \UnaryInfC{$\SequentCLJ{\psi \impl \varphi}$}
      \bottomAlignProof
      \DisplayProof
      \quad
      \AxiomC{$\SequentCLJ(\ThAssumsEnv, \psi){\varphi}$}
      \AxiomC{$\SequentCLJ(\ThAssumsEnv)[\GuardedAssumsEnv , \CoindAssumsEnv][\emptyset]{\xi}$}
      \RightLabel{\LJImplLT}
      \BinaryInfC{$\SequentCLJ(\ThAssumsEnv, \xi \impl \psi){\varphi}$}
      \bottomAlignProof
      \DisplayProof
      \\
      \AxiomC{$\SequentCLJ[\GuardedAssumsEnv, \psi]{\varphi}$}
      \AxiomC{$\SequentCLJ{\xi}$}
      \RightLabel{\LJImplLG}
      \BinaryInfC{$\SequentCLJ[\GuardedAssumsEnv, \xi \impl \psi]{\varphi}$}
      \bottomAlignProof
      \DisplayProof
      \\
      \AxiomC{$\SequentCLJ[\GuardedAssumsEnv][\CoindAssumsEnv, \varphi]{\varphi}$}
      \RightLabel{\LJCoFixRule}
      \UnaryInfC{$\SequentCLJ{\varphi}$}
      \bottomAlignProof
      \DisplayProof
      \quad
      \AxiomC{$\SequentCLJ{\psi}$}
      \AxiomC{$\SequentCLJ[\GuardedAssumsEnv, \psi]{\varphi}$}
      \RightLabel{\LJCut}
      \BinaryInfC{$\SequentCLJ{\varphi}$}
      \bottomAlignProof
      \DisplayProof
    \end{empheq}
  \end{spreadlines}}
  \vspace*{-1em}
    \caption{\emph{\footnotesize{The rules for \CLJ{}, standard structural rules are assumed.}}}
    \label{fig:rules-CLJ}
    \vspace*{-1.5em}
  \end{figure}

\begin{theorem}[Cut is not eliminable in \CLJ{}]
  Any proof of $\SequentCLJ(\{\all{x} p(f(x)) \impl p(x)\})[\emptyset][\emptyset]{p(a)}$
  uses the $\LJCut$ rule.
\end{theorem}
\emph{Proof.}
  To prove the theorem we will construct a set   $\mathcal{S}$ of \emph{bad sequents} in a proof tree for $\SequentCLJ(\{\all{x} p(f(x)) \impl p(x)\})[\emptyset][\emptyset]{p(a)}$,
  such that the following conditions hold:
\begin{enumerate}
\item The rule {\footnotesize{$\LJAxiom$}} does not belong to $\mathcal{S}$;
\item For every instance of any rule except {\footnotesize{\LJCut}}, if the conclusion belongs to $\mathcal{S}$ then at least one premise belongs to $\mathcal{S}$;
\item Sequent $\SequentCLJ(\{\all{x} p(f(x)) \impl p(x)\})[\emptyset][\emptyset]{p(a)}$ belongs to $\mathcal{S}$.
\end{enumerate}

If these three conditions hold, then there are no finite proofs without cut for any sequent in $\mathcal{S}$, including the sequent from the theorem statement.

Let us now construct $\mathcal{S}$. It consists of sequents of the form $\SequentCLJ{p(N)}$ (with an arbitrary term $N \in \FOSimpleTerms{\TSig}$) such that:
\begin{itemize}
\item \footnotesize{$\ThAssumsEnv \subseteq \{p(t) \mid t \in \FOSimpleTerms{\TSig} \colon t \neq f^i(N) \;\; \forall i \ge 0 \} \cup \{\all{x} p(f(x)) \impl p(x)\} \cup \{p(f(t)) \impl p(t) \mid t \in \FOSimpleTerms{\TSig} \},$}
\item \footnotesize{$\GuardedAssumsEnv \subseteq \{p(t) \mid t \in \FOSimpleTerms{\TSig} \colon t \neq f^i(N) \;\;  \forall i \ge 0 \},$}
\item \footnotesize{$\CoindAssumsEnv \subseteq \{p(t) \mid t \in \FOSimpleTerms{\TSig} \colon t \neq f^i(N) \;\; \forall i > 0 \}.$}
\end{itemize}
So, we allow in premises only formulae of the form $p(t)$ with $t$ different from $N$ with $f$ applied any number of times, we also allow succedent in the set of unguarded premises (note $>$ instead of $\ge$ there) and the given clause  $\all{x} p(f(x)) \impl p(x)$ in the set of theory assumptions (uninstatiated or instantiated with an arbitrary term).

We now only need to check that the conditions for a set of \emph{bad sequents} hold.

(1) Obvious, as we explicitly forbade the succedent from the guarded assumptions.

(2) There are very few rules except  {\footnotesize{\LJCut}} that we can apply to a sequent of this form.
We can apply {\footnotesize{\LJAllLT}}, {\footnotesize{\LJCoFixRule}} or the structural rules,
which will keep us in $\mathcal{S}$ simply by its definition.
The only non-trivial case is if we apply the {\footnotesize{\LJImplLT}}-rule to use an assumption $p(f(M)) \impl p(M)$ with some term $M$. We will consider two subcases here:

(2.1) $M \neq f^i(N)$ for all $i \ge 0$. Then the premise \[\SequentCLJ(\ThAssumsEnv, p(M)){p(N)}\] belongs to $\mathcal{S}$, as in this subcase $p(M)$ satisfies the condition for assumptions from $\ThAssumsEnv$.

(2.2) $M = f^k(N)$ for some $k \ge 0$. Then we can show that the other premise \[\SequentCLJ[\GuardedAssumsEnv, \CoindAssumsEnv][\emptyset]{p(f(M))}\] belongs to $\mathcal{S}$. We can rewrite it as \[\SequentCLJ[\GuardedAssumsEnv,\CoindAssumsEnv][\emptyset]{p(f^{k+1}(N))}.\] As all assumptions of the form $p(t)$ safisfy $t \neq f^i(N) \;\; \forall i > 0$, because the conclusion belongs to $\mathcal{S}$, they therefore satisfy $t \neq f^{i + k + 1}(N) \;\; \forall i \ge 0$.

(3) Obvious. \hfill $\Box$

Note that, because of its simplicity, this result will be replicable in many sequent
calculi like, for instance, the classical system \LK{}~\cite{Troelstra00:BasicProofTheory,SU06}.

\section{CoLP Derivations as Cut-free Proofs}\label{sec:exist}

Intuitively, the loop detection method of
CoLP~\cite{GuptaBMSM07} amounts to finding atoms $A$ and $B$ in an SLD-derivation such that $A$ and $B$ unify.
This, possibly circular, unifier gives rise to a possibly infinite
atom given by a \emph{rational tree}~\cite{Courcelle83}.
It may seem plausible to conjecture that CoLP's set of all provable atoms corresponds to the
set of all rational trees in the program's model, but this conjecture is disproven by our example of
the logic program $\Gamma_T$ and the goal $p(a)$, that can be represented by a rational tree, but
cannot be proven in CoLP.
This section proposes an alternative characterisation of provability in CoLP as a set of atoms provable in cut-free \CLJ{}.
Providing a different perspective on this result, \citeN{DAZ20} have recently shown that CoLP covers all regular infinite SLD-trees.
The regular proofs of \citeN{DAZ20} correspond to finite cut-free \CLJ{} proofs in which the coinduction hypothesis/goal encapsulates the structure of the entire infinite regular proof.

To establish our result, we need to allow first-order guarded fixpoint terms
in goals and in (coinductive) assumptions in $\Gamma_T$, $\Gamma_A$ and $\Gamma_C$.
The main technical idea of this section is to show how circular unifiers of CoLP convert into first-order fixpoint terms.
This conversion delivers us the theoretical result we seek, and may also open the way for using CoLP within richer coinductive theorem provers.

\emph{Substitution} $\sigma$ is a finitely supported function from variables to simple first-order terms (i.e. terms in $\FOSimpleTerms{\Sigma}$).
As usual, a substitution $\sigma$ can be  extended to a function from $\FOSimpleTerms{\Sigma}$ to $\FOSimpleTerms{\Sigma}$
 by taking $(f \ t_1 \ldots t_n)[\sigma] = f \ t_1[\sigma] \ldots t_n[\sigma] $,
 whenever $f$ is a constant in $\Sigma$.
If $\sigma_1$ and $\sigma_2$ are substitutions, then their \emph{composition}
$\sigma_1 \circ \sigma_2$ is defined by
$(\sigma_1 \circ \sigma_2)(x) = \sigma_2(x)[\sigma_{1}]$.
A substitution $\sigma$ is a \emph{unifier} for $t, u \in \FOSimpleTerms{\Sigma}$,
if $t[\sigma] = u[\sigma]$, it is a \emph{matcher} if $t[\sigma] = u$.
%
We say a substitution $\sigma = \subst{t/x}$
is \emph{circular} if $x$ appears among the free variables of $t$.
For example, $\subst{{\scons(0, x)} / x}$ is a circular substitution.

In order to represent circular substitutions as fixpoint terms,
we need to extend the notion of substitution to \emph{fix-substitution},
which is defined as a finitely supported function from
variables to guarded first-order terms, i.e. terms in  $\GuardedFOTerms{\TSig}$.
We will denote fix-substitutions by $\delta, \delta_{0}, \delta_{1}, \dotsc$ to
distinguish them from simple first-order substitutions.
Fix-substitutions extend to functions
$\GuardedFOTerms{\TSig} \to  \GuardedFOTerms{\TSig}$ by capture-avoiding substitution.

A fix-substitution $\delta$ is a \emph{fixpoint unifier}
for $t, u \in  \GuardedFOTerms{\TSig}$, if $t[\delta] \conv u[\delta]$, where we recall $\conv$
to be conversion with $\fix$- and $\beta$-reduction (see~\cite{BasoldKL19-2}).

We first show that, given a circular substitution $\sigma = \ssubst{f \, \vec{t}}{x}$, we can obtain a
fix-substitution $\delta = \ssubst*{\fix[x] f \, \vec{t}}{x}$.
For example, the circular substitution
$\subst{{\scons(0, x)} / x}$  gives rise to the fix-substitution
$[\fix[x] \scons(0, x) / x]$.
Finding such substitutions in the general case requires some additional
machinery, as the following example shows.

\begin{example}[Circular substitutions do not result in circular unifiers]
  \label{ex:co}
  For the two atoms $p(f(x,y),g (x, y))$ and $p(x, y)$,
  let $\sigma_1 = [f(x, y) / x]$ and
  $\sigma_2 = [g(x, y) / y]$.
  We would like to define a unifier by $\sigma = \sigma_2 \comp \sigma_1$.
  However, the composition will result in
  $\sigma = [f(x, g(x, y))  /  x, \, g(x, y) / y]$,
  which is not quite the unifier $[f(x, y) / x, \,  g(x, y) / y]$ that we expect.
  For this reason, the circular substitutions are not composed in CoLP, but are
  simply taken as sets of equations, like
  $\set{x = f(x, y) \, , y = g(x, y)}$.
\end{example}

\noindent We need a notion of composition for circular
substitutions, in order to have proper circular unifiers as part of the language.
And this is where we make use of fixpoint terms.

\begin{definition}[Unifying equations]
  Given $t, u \in  \FOSimpleTerms{\TSig}$, a set $\mathcal{U}_{t,u}$ of
  \emph{unifying equations} is defined inductively as follows:
  \begin{enumerate}
  \item if $t = x$ for some $x \in \Vars$, then $\mathcal{U}_{t,u} = \set{x = u}$,
  \item if $u = x$ for some $x \in \Vars$, then $\mathcal{U}_{t,u} = \set{x = t}$,
  \item if $t = f \, t_1 \cdots \, t_n$ and $u = f \, u_1 \cdots \, u_n$,
    then $\mathcal{U}_{t,u} = \bigcup_{k = 1}^{n} \mathcal{U}_{t_k, u_k}$, and
  \item $\mathcal{U}_{t,u} = \emptyset$  otherwise.
  \end{enumerate}
  Two simple first-order atoms $A = p \, t_1 \cdots \, t_m$ and $B = p \, u_1 \cdots \, u_m$ have as set of unifying equations $\mathcal{U}_{A,B} = \bigcup_{k = 1}^{m} \mathcal{U}_{t_k, u_k}$.
\end{definition}
Clearly, if $\mathcal{U}_{t,u}$ is empty,
then $t$ and $u$ are not unifiable.
If the set of unifying equations
contains at most one equation for each variable,
we say that it is \emph{linear unifying}.

The mentioned set $\set{ x = f(x, y), y = g(x, y)}$ is
linear unifying for  $p(f(x, y), g(x, y))$ and $p(x, y)$.
We refer an interested reader to~\cite{Courcelle83,GuptaBMSM07} for a more
detailed study of properties of unifying equations.
Notably, every system of such equations has the most general unifier that is
rational.

\begin{definition}[Circular Unifier]\label{def:circ}
  Let $A, B \in \foSimpleAt$ have a set of linear unifying equations
  $\mathcal{U}_{A,B} = \bigcup_{i=1}^n \{ x_i = t_i \}$.
  We can define a sequence of fix-substitutions
  $\delta_0, \delta_1, \ldots, \delta_n$, such that $\delta_k$ unifies the
  first $k$ equations, as follows:
  \begin{align*}
    \SwapAboveDisplaySkip
    \delta_0 &= \id \\
    \delta_{i+1} &=
    \begin{cases}
      \delta_i,
      & \text{if } t_{i+1}[\delta_i] = x_{i+1} \\
      [t_{i+1}[\delta_i] /\ x_{i+1} ] \comp \delta_i,
      & \text{if } x_{i+1} \not\in FV(t_{i+1}[\delta_i])  \\
      [\fix[x_{i+1}] t_{i+1}[\delta_i] /\ x_{i+1} ] \comp \delta_i,
      & \text{if } t_{i+1}[\delta_i] \ne x_{i+1}, \, x_{i+1} \in FV(t_{i+1}[\delta_i])
    \end{cases}
  \end{align*}

  Then the fix-substitution $\delta_n$ is called the \emph{circular unifier}
  for $A$ and $B$.
\end{definition}

\begin{example}[Circular Unifiers]\label{ex:com}
  Given the set $\mathcal{U} = \set{ x = f(y), y = g(x) }$ for the atoms
  $p(f(y), (g(x))$ and $p(x, y)$,
  the circular unifier will be
  $\delta = [ \fix[y] g(f(y))/ y ] \comp [ f(y)/ x ]$,
  which amounts to\\ $[ f(\fix[y] g \, (f(y))) / x, \fix[y] g(f(y)) / y]$.

  We continue with the equations
  $\set{ x = f(x,y), y = g(x,y)}$, and atoms
  $p \, (f(x, y),g(x, y))$ and $p(x, y)$
  from \Cref{ex:co}.
  From \Cref{def:circ}, we obtain the desired circular unifier
  $[ \fix[y] g \, (\fix[z] f(z, y), y) / y ]
  \comp [ \fix[x] f(x, y) /x ]$,
  which in turn is equal to the substitution
  $[ \fix[x] f(x, \fix[y] g(\fix[z] f(z, y), y)) /\ x, \fix[y] g(\fix[z] f(z, y), y) /\ y]$.
\end{example}

The following lemma shows that circular unifiers are
fixpoint unifiers.

\begin{lemma}[Circular unifier is a fixpoint unifier]
  \label{lem:circ2}
  Let  $A, B \in \foSimpleAt$ and let $\sigma$ be their circular unifier.
  Then, $A[\sigma] \conv B[\sigma]$.
\end{lemma}

We  can now use circular unifiers to generate coinduction hypotheses.

\begin{example}[Coinduction Hypothesis from Circular Unifiers]
  Taking $P_{\stream 0}$
  and the goal $\stream \, (\scons(0, x'))$,
  CoLP finds $\set{ x' = \scons(0, x')}$ as circular unifier. This corresponds to
  the coinduction hypothesis $\stream \, (\fix[x] \scons(0, x))$.
\end{example}

\citeN{SimonEtAl06} have shown that the method of loop detection is sound relative to the complete Herbrand models of logic programs.
CUP, a cut-free fragment of \CLJ{} was also shown to be sound relative to the complete Herbrand models\cite{BasoldKL19-2}.
We only need to show that we form fixpoint terms from loops correctly.

\begin{theorem}[CoLP proofs in cut-free \CLJ{}]
  Let $\Gamma_T$ be a logic program and $A \in \foSimpleAt$.
  If CoLP returns a proof and a circular substitution $\theta$ for $\Gamma_T$ and $A$ that is given
  by a set $\mathcal{U}$ of linear unifying equations, then:
  \begin{itemize}
  \item there exists a circular unifier $\delta$ for $\mathcal{U}$,
  \item and there is a cut-free proof for  $\SequentCLJ[\emptyset][\emptyset]{\exist{\vec{x}} A}$.
  \end{itemize}
\end{theorem}
  \emph{Proof.} The first property follows from the construction of \Cref{def:circ} and
  \Cref{lem:circ2}.
  The second property is also proven constructively, by constricting a \CLJ{} proof in which, as the first step, the
  existential variables $\vec{x}$ are substituted as in $\delta$, and then  the proof for
  $\SequentCLJ[\emptyset][\emptyset]{A[\delta]}$ proceeds  by  {\footnotesize{$\LJCoFixRule$}}, taking $A[\delta]$ as coinduction hypothesis. The proof is completed by
  following the same resolution steps (emulated by a combination of {\footnotesize{$\LJAllLT$}},  {\footnotesize{$\LJConjLT$}},  {\footnotesize{$\LJImplLT$}},  {\footnotesize{$\LJAxiom$}}) as in the given CoLP derivation, applying the coinduction hypothesis where loop detection was applied by CoLP (using {\footnotesize{$\LJAxiom$}}).
  \hfill $\Box$

Taking, for example, the logic program $P_{\stream 0}$ and the input formula $\stream \, x$,
and having obtained $\stream\, (\fix[x] \scons(0,x))$ from CoLP's circular unifier,
we will be able to prove
$P_{\stream 0} + \emptyset + \emptyset \vdash \stream \, ( \fix[x] \scons(0,x))$ by coinduction.

We provide implementation of the method of turning CoLP-style circular unifiers into \CLJ{} (or CUP) proofs\cref{fn:impl}.

\section{Coinductive Theory Exploration and  Implementation}\label{sec:irreg}

Coinductive proofs in first-order logic are, in general, not recursively enumerable.
We thus have to resort to smaller, cut-free, fragments of coinductive theories,
as in CoLP or CUP, for automated proving.
As a consequence, we can only hope for heuristics to find suitable cut formulae (and coinduction hypotheses) in the general case.

We present here a new method of \emph{coinductive theory exploration} for \CLJ{}, and provide its
implementation.\cref{fn:impl}
We automate cut-free proof search in \CLJ{} (equivalently in CUP). That is, given a logic program $P$ and a goal $G$, we can (semi)decide whether
$\SequentCLJ(P)[\emptyset][\emptyset]{G}$ holds.
If the automated search fails,
a theory exploration method is invoked. It analyses proof-patterns and in particular loops that arose in the failed proof of $G$.
It generalises this information in a form of a candidate
coinduction hypothesis $CH$. The tool then tries to prove $\SequentCLJ(P)[\emptyset][\emptyset]{CH}$ by coinduction. If the proof fails, $CH$ is discarded.
If the proof succeeds, the tool re-attempts to prove $\SequentCLJ(P)[CH][\emptyset]{G}$.

Our implementation incorporates three kinds of methods. Firstly, we benefit from CoLP's method of searching for circular unifiers, whenever such exist.
Secondly, we implement the method of \citeN{FKS15} that worked for cases when $CH$ was limited to H-formulae (without fixpoint or $\lambda$-terms).
Finally, we implement a completely novel heuristic that covers the case when $CH$ is a G-formula with (guarded) higher-order fixpoint terms.
This method is restricted to logic programs that define non-periodic streams, such as  $P_{\fromP}$
or the program that defines the stream of Fibonacci numbers in \Cref{ex:fib}. However, our implementation is done in a modular way and will admit novel heuristics and extensions in the future.

\begin{figure}[t]
  \makebox[\textwidth]{
\begin{tikzpicture}[every tree node/.style={},
   level distance=1.20cm,sibling distance=1.00cm,
   edge from parent path={(\tikzparentnode) -- (\tikzchildnode)}, font=\scriptsize]
\Tree [.\node{\mybox{$\stream(x)$}};
         ]
       \end{tikzpicture}
  $\stackrel{\scons(0, x')/ x}{\leadsto}$

\begin{tikzpicture}[every tree node/.style={},
   level distance=1.20cm,sibling distance=1.00cm,
   edge from parent path={(\tikzparentnode) -- (\tikzchildnode)}, font=\scriptsize]
\Tree [.\node{\mybox{$\stream \, (\scons(0,  x'))$}};
         [.\node{$\stream(x') $};
         ]
         ]
       \end{tikzpicture}

      $\stackrel{\scons(0, x'')/ x'}{\leadsto}$
\begin{tikzpicture}[every tree node/.style={},
   level distance=1.20cm,sibling distance=1.00cm,
   edge from parent path={(\tikzparentnode) -- (\tikzchildnode)}, font=\scriptsize]

\Tree [.\node{\mybox{$\stream \, (\scons( 0, \scons(0,  x'')))$}};
        [.\node{$\stream (\scons(0, x'')) $};
         [.\node{$\stream(x'') $};
         ]
         ]]
       \end{tikzpicture}
     }
     \vspace*{-0.5em}
   \caption{\emph{\footnotesize{Rewriting tree transitions for $P_{\stream 0}$.  Boxes show tree roots.}}}
  \label{fig:stream}
   \vspace*{-1.5em}
 \end{figure}
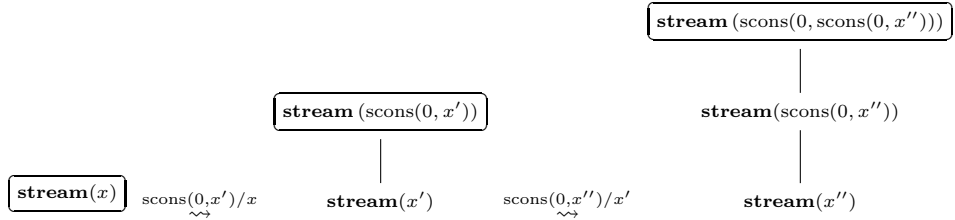

From the technical point of view, our implementation benefits from using S-resolution by~\citeN{KL17} instead of SLD-resolution, when it comes to exploring recursive proof patterns.
S-resolution helps to separate out the term-matching and unification components of
computations, by doing term-matching steps eagerly, and unification steps lazily.
\Cref{fig:stream} shows term-matching steps as vertical transitions and unification steps as
horizontal transitions.
Each vertical block, also called a \emph{rewriting tree}, shows clearly reductions of the stream
constructor.
This is a useful property, as it helps to see the relation between the constructor and other
arguments.

Formally, a \emph{rewriting tree} is defined by a map from a tree domain to  $\foSimpleAt$. For definitions of infinite trees
as maps from infinite tree domains see e.g.~\cite{Courcelle83}.
We write $\omega$ for the set of non-negative integers and $\omega^*$ for the
set of all finite lists over $\omega$.
Lists are denoted by $(i,\ldots,j)$ where $i,\ldots,j\in\omega$.
The empty list is denoted $\epsilon$.
If $u,v\in\omega^*$, then $(u,v) \in \omega^{*}$ is the concatenation of $u$
and $v$.
If $u\in\omega^*$ and $i \in \omega$, then $(u,i)$ denotes the list $(u,(i))$.
Finally, $u > v$ if $u = (v , v')$ for some non-empty $v'$.
%
%
  A set $L \subseteq \omega^*$ is a \emph{(finitely branching)
    tree domain} provided: \vspace*{-0.5em}
  \begin{itemize}
  \item $\all{u \in \omega^*} \all{j \in \omega}$
    if $(u,j) \in L$ then $u \in L$ and $\all{i<j} (u,i) \in L$; and
  \item the set $\setDef{i\in \omega}{(u,i)\in L}$ is finite for all $u \in L$.
  \end{itemize} \vspace*{-0.5em}
A non-empty tree domain always contains $\rewTreeRoot$,
      which we call its {\em root}.

          \begin{definition}[Rewriting tree]
A rewriting tree for  $A \in \foSimpleAt$ and a logic program $P$ is
a map $T \from L \to \foSimpleAt$  satisfying:
\begin{itemize}
\item $T(\rewTreeRoot) =  A$, and
\item $(u, i) \in L$
  and $T(u , i) = B_i[\sigma]$,
  if there is
  $(\all{\vec{x}} B_1\conj \dotsm \conj B_i \conj \dotsm \conj B_n \impl B) \in P$ and
  $T(u) =  B[\sigma]$.
  If $n = 0$, we write $T(u, i) = \Box$.
\end{itemize}
\end{definition}

\noindent In the above definition, we assume the standard method of
\emph{renaming variables apart} used to avoid circular
unification.

\begin{figure}[t]
  \makebox[\textwidth]{

\begin{tikzpicture}[every tree node/.style={},
   level distance=1.20cm,sibling distance=1.00cm,
   edge from parent path={(\tikzparentnode) -- (\tikzchildnode)}, font=\scriptsize]
\Tree [.\node{\mybox{$\fromP(0, y)$}};
         ]
       \end{tikzpicture}
  $\stackrel{\scons(0, y')/ y}{\leadsto}$
   \begin{tikzpicture}[every tree node/.style={},
   level distance=1.20cm,sibling distance=1.00cm,
   edge from parent path={(\tikzparentnode) -- (\tikzchildnode)}, font=\scriptsize]
\Tree [.\node{\mybox{$\underline{\fromP(0,\scons(0,  y'))}$}};
         [.\node{$\underline{\fromP(s(0), y') }$};
         ]
         ]
       \end{tikzpicture}

\begin{tikzpicture}[every tree node/.style={},
   level distance=1.20cm,sibling distance=1.00cm,
   edge from parent path={(\tikzparentnode) -- (\tikzchildnode)}, font=\scriptsize]
\Tree [.\node{\mybox{$\fromP(x, z)$}};
         ]
       \end{tikzpicture}
  $\stackrel{\scons(x,z')/ z}{\leadsto}$
\begin{tikzpicture}[every tree node/.style={},
   level distance=1.20cm,sibling distance=1.00cm,
   edge from parent path={(\tikzparentnode) -- (\tikzchildnode)}, font=\scriptsize]
\Tree [.\node{\mybox{$\underline{\fromP(x,\scons(x,  z'))}$}};
         [.\node{$\underline{\fromP(s(x),z')}$};
         ]
         ]
\end{tikzpicture}
}\vspace*{-0.5em}
\caption{\emph{\footnotesize
    Left: Transition of two rewriting trees for the goal formula
    $\fromP(0, y)$. Underlined are the critical pairs.
    Right: abstract representation of the second tree on the left, and a transition for the abstract tree.}}
\label{fig:ex2}
\vspace*{-1.5em}
\end{figure}
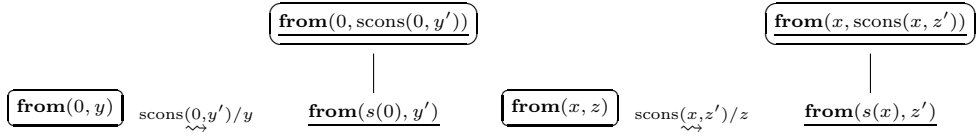

Given the rewriting tree $T$ (for $P$ and $A \in \foSimpleAt$),
such that some leaf $T(u)$ unifies with the head of a clause in $P$ via a substitution $\theta$,
we can construct a rewriting tree $T_1$ for $P$ and $A[\theta]$.  We write $T \rewTreeTran   T_1$ to denote this tree transition.
\Cref{fig:stream,fig:ex2} show such transitions.
We say that a logic program is \emph{productive}~\cite{KL17} if it admits only finite
rewriting trees, thus requiring tree transitions for any infinite computation.
$P_{\stream0}$ and $P_{\fromP}$ are productive programs, whereas $\Gamma_T$ is not.
For the rest of this section, we will be working only with productive programs (as all stream definitions give rise to such).
Our implementation\cref{fn:impl} also covers coinductive theory exploration for infinite rewriting trees, following the method of \citeN{FKS15}.

The new heuristic for programs defining non-periodic streams is based on three ideas:

\paragraph{Idea 1: Non-periodic streams can be described by higher-order fixpoint terms.}
Usually, definitions of non-periodic streams rely on iterating some function that modifies its
arguments recursively, and thus computes the stream members that do not unify among each other.
In the case of $P_{\fromP}$, the map $s$ modifies, say, $0$ to $s(0)$, $s(s(0))$, and so on.
Thus, definitions of such streams involve construction of a fixpoint of a function, rather than of a term variable.
We explore this connection between non-periodic stream patterns and  higher-order recursive functions.

We assume for the remainder of this section that
the goal of our  proof is an atom $A\in  \foSimpleAt$ that is built of a predicate that defines some
infinite stream, that is,
\begin{equation*}
  A = p_{\stream} \, t^{in}_1 \, \dotsm \, t^{in}_j \,  x^{out} \, ,
\end{equation*}
and the program that defines  $p_{\stream}$ is productive.
Moreover, $x^{out}$  is the \emph{output} argument in the process of computation of streams,
the terms $t^{in}_1, \dotsc, t^{in}_j$ contain no variables and provide the inputs for the stream construction.
For example, in the goal $\fromP(0,y)$, $0$ is the input and $y$ is the output.

We thus exclude programs like $P_{double}$:
\begin{equation*}
  \kappa_{double} :
  \all{x\,y\,z_{1}\,z_{2}} double(s(x), s(s(y)), z_{1}, z_{2})
  \impl double(x, y, \scons(x, z_{1}), \scons(y, z_{2}))
\end{equation*}
that defines two streams of numbers.
This restriction is made in order to reduce the notational clutter.
The method we present should generalise well to these cases, modulo keeping track of term positions.

Finally, we require that all clauses in the given program are linear, that is, contain at most one
recursive call (all examples given so far are linear).

\begin{definition}[Higher-order fixpoint stream definition]\label{df:df}
  Given a logic program $P$, and an $n$-ary predicate $p_{\stream}$ in $P$ that defines a stream
  $s$ with the function (stream constructor) $\scons$ in its last argument,
  we say $s^{fix}$ given by
  $\fix[f] \lam{x_1 \, \dotsm \, x_{n-1}} \scons\, x_1 \,  (f \, t^?_1 \, \dotsm \, t^?_{n-1})$ is a
  \emph{higher-order fixpoint definition of $s$} if there exist
  $t^?_1, \ldots , t^?_{n-1}   \in \foSimpleAt $ such that
  \begin{equation*}
    \SequentCLJ(P)[\emptyset][\emptyset]{
      \all{x_1 \, \dotsm \, x_{n-1}} p_{\stream} \, x_1 \, \dotsm \, x_{n-1} \, s^{fix}}.
  \end{equation*}
  In this case we call
  $\all{x_1 \, \dotsm \, x_{n-1}} p_{\stream} \, x_1 \, \dotsm \, x_{n-1} \, s^{fix}$ the
  \emph{candidate coinduction hypothesis} for $P$ and  $p_{\stream}$.
\end{definition}

We can now see that coinductive theory exploration for higher-order fixpoint stream definitions
amounts to  search for  suitable   $t^?_1, \ldots , t^?_{n-1}   \in  \FOSimpleTerms{\TSig}$;
these terms contain the functions that will be iterated by $\fix$.
We next define a possible heuristic for this search.

\paragraph{Idea 2: Resolution by term matching helps to find and analyse irregular recursive proof patterns.}
This idea has been explored in detail by~\citeN{FKS15} in the context of infinite rewriting trees.
 We follow that line of work and use the Paterson condition to find irregular recursive patterns in rewriting trees:

\begin{definition}[Paterson Condition \cite{SulzmannDJS07}]
  Let $\FunSym{A}$, $\FVar{A}$  denote the
  multiset of term symbols and the multiset of free variables in $A$.
  The \emph{Paterson condition} is satisfied by an H-formula
  $\all{\vec{x}}( B_1 \conj \dotsm \conj B_n \impl A)$
  if $(\FunSym{B_i} \cup \FVar{B_i})
  \subset (\FunSym{A} \cup \FVar{A})$ for each $B_i$.
  The pair of simple first-order atoms $\langle A, B \rangle$ is called a \emph{critical pair}, if $\all{\vec{x}} B \impl A$ does not satisfy the Paterson condition.
\end{definition}

Irregular proof traces usually give rise to critical pairs. To use this fact, we say
a rewriting tree $T$ is an \emph{irregular rewriting tree} if, each leaf $T(u)$ is either a $\Box$ or  forms a critical pair $\langle T(\epsilon), T(u) \rangle$ with the root $T(\epsilon)$.
In \Cref{fig:ex2}, the second tree is an irregular rewriting tree, but \Cref{fig:stream} has none.

\paragraph{Idea 3: We need anti-unification to turn irregular recursive patterns into higher-order fixpoints.}
As  \Cref{fig:ex2} shows, just having irregular rewriting trees does not solve the problem of finding higher-order stream definitions.
Given a sequence of rewriting tree transitions, we need to be able to abstract from concrete constants to general recursive patterns.
We implement our own version of the algorithm of anti-unification by  \cite{plotkin1970note} to obtain abstract representations of rewriting trees.

  Let $M, N \in \FOSimpleTerms{\TSig}$ be two simple first-order terms,
  possibly with free variables.
  We write $M \leq N$ if there is a substitution $\sigma$, such that
  $M[\sigma] = N$.
  A term $A$ is a \emph{generalisation} of $M$ and $N$, if
  $A \leq M$ and $A \leq N$.
The following lemma shows that the order and term generalisation are sensible:

\begin{lemma}
  \label{lem:terms-poset-generalisation-filtered}
  The order $\leq$ makes $ \FOSimpleTerms{\TSig}$ a poset.
  Moreover, for any two terms $M$ and $N$, the set
  $\setDef{A \in  \FOSimpleTerms{\TSig}}{A \text{ generalises } M \text{ and } N}$
  is filtered, that is, for all generalisations $A$ and $B$ there is a
  generalisation $C$ with $A \leq C$ and $B \leq C$.
\end{lemma}

\noindent Since the set of generalisations is filtered and  bounded, there
is a maximal generalisation.

\begin{definition}[Anti-Unifier \cite{plotkin1970note}]
  The \emph{anti-unifier} of two terms $M$ and $N$
  is the maximal (or least general) generalisation of $M$ and $N$,
  and will be denoted by $M \sqcap N$.
  This extends in the obvious way to the anti-unifier of atoms.
\end{definition}

For example, $p(a) \sqcap p(b) = p(x)$.

\begin{definition}[Abstract Representation of a Rewriting Tree]
  \label{abs:rep}
Let $T$ be a rewriting tree.  Suppose that $\pair{T(\epsilon), T(v_1)},\dotsc, \pair{T(\epsilon), T(v_n)}$ are all critical pairs, where  $T(v_1), \ldots , T(v_n)$   are leaves of $T$.
  Let us define $A \in \foSimpleAt$ to be the anti-unifier $T(\epsilon) \sqcap \parens*{\bigsqcap_{i=1}^{n} T(v_{i})}$.
  The \emph{abstract representation} $T'$ of  $T$ is defined as:
  \begin{itemize}
  \item $T'(\epsilon) = A$
  \item $T'(u, i) =B_i[ \sigma ]$  if $T'(u) = B[ \sigma ]$ and $(B_1, ..., B_n \rightarrow B) \in
P$. When $n = 0$, we write $T'(u , i) = \Box$.
  \item $T'(u)$ is undefined if $u > v_i$ for some $T(v_i)$ ($1 \leq i \leq n$), i.e.  $T'(v_1), \ldots , T'(v_n)$   are leaves of $T'$.
  \end{itemize}
\vspace*{-1em}
\end{definition}

It is easy to see that there exists an abstract representation for each  irregular
rewriting tree.
In \Cref{fig:ex2} the third tree is the abstract representation of the second tree.
It abstracts away from concrete terms to more general recursive patterns.
However, it is really the fourth tree obtained by transition from the third tree that is of interest.
We formalise the above intuition as follows.
When a proof search
\begin{itemize}
\item starts with a program $P$, a goal $A$ and a rewriting tree $T$ for $P$ and $A$,
\item finds an irregular rewriting tree $T'$ and
  its corresponding abstract tree $T''$
\item and then proceeds constructing tree transitions from $T''$,
\end{itemize}
we will say the search
is done in an \emph{abstract search domain for $P$ and $A$}.
\Cref{fig:ex2} shows  rewriting trees in an abstract search domain
for $P_{\fromP}$ and $\fromP(0, y)$.

The next definition uses Ideas 1, 2, 3 to formulate the novel method of search for higher-order fixpoint terms that capture irregular proof patterns:

\begin{definition}[Heuristic Search for Coinduction Hypotheses for Irregular Streams]\label{def:search}
  \label{def:fxirr}
  Let $T$ be an irregular rewriting tree  in an abstract search domain for $P$, with the
  root $A =  p \, x_1 \, \dotsm \, x_{n-1} \, t_n$,
  where $p$ defines a stream $s$.
  Let $A$ and a leaf  $T(v) =  p \, t_1 \, \dotsm \, t_{n-1} \, x_n$ form a critical pair.
  Then a candidate higher-order fixpoint definition of $s$ is obtained by taking
   $t^?_i = t_i$ (as in \Cref{df:df}).
\end{definition}

\begin{example}[Candidate Coinduction Hypothesis]

The final tree in  \Cref{fig:ex2} gives rise to a critical pair.
Applying \Cref{df:df,def:fxirr},   we obtain the candidate fixpoint term
{\small{$f_{\fromP} := \fix[f] \lambda\, x. \scons(x, f (s(x))$}} and the candidate
coinduction hypothesis {\small{$\all{x} \fromP(x, f_{\fromP})$}}.
Then we prove
{ \small{$P_{\fromP} + \emptyset + \emptyset \vdash \all{x} \fromP(x, f_{\fromP})$}}.
The original goal, $\exist{z} \fromP(0, z)$ is then obtained by an application of {\small\LJCut{}}, i.e. we prove {\small{$P_{\fromP} + (\all{x} \fromP(x, f_{\fromP}) + \emptyset \vdash \exist{z} \fromP(0, z)$}} by instantiating
$z$ with $f_{\fromP} \, 0$.

\end{example}

\begin{example}[More Complex Coinduction Hypotheses]\label{ex:fib}
  Taking the program $P_{fib}$ computing pseudo-Fibonacci sequence
  \begin{equation*}
    \kappa_{fib} :   \all{xy} fib(y, x+y,  z) \impl fib(x, y, \scons(x,  z))
  \end{equation*}
  and a goal $\exist{z} fib(0, 1, z)$, we obtain an abstract representation of a rewriting tree with the root $fib(x, y, \scons(x,  z))$, and the leaf
  $fib(y, (x+y), z)$. The  corresponding
  candidate stream definition $f_{fib}$ is given by
  $\fix[f] \lam{x \, y} \scons(x, f(y, x+y))$, and the coinduction hypothesis is
  $\all{x \, y} fib(x, y, f_{fib})$.
\end{example}

\section{Conclusions, Related and Future Work}
\label{sec:concl}

This paper contributes to previous attempts to give proof-theoretic and constructive interpretation to logic and answer-set programming:~\cite{MNPS91,MN12,FK16,SchubertU18,BasoldKL19-2}.
Here, our goal was two-fold. Firstly, we showed that cut is not eliminable in a coinductive first-order sequent calculus.
Secondly, we analysed the current state of the art in coinductive logic programming (given by CoLP) in the proof-theoretic terms,
exposing that CoLP derivations in fact correspond to cut-free proofs in \CLJ{}. Both of these results led to a conclusion that
any further progress in coinductive logic programming is only possible by introducing richer heuristics of coinductive theory exploration.
With this in mind, we proposed a composite method, similar to the famous \emph{Boyer-Moore Waterfall Model}~\cite{Boyer79:ComputationalLogic}, which incorporates automated proofs in \CLJ{}, as well as several existing and one novel heuristics searching for suitable coinduction hypotheses.
We provided a prototype implementation.\cref{fn:impl}

The novel theory exploration heuristic that we provided serves mainly as an illustration of the range of methods (S-resolution, anti-unification, higher-order fixpoint terms)
that can be employed in the future for
a systematic synthesis of coinduction hypotheses for proofs in Horn clause theories. We hope to investigate further extensions in the future. 

Coinduction is now implemented in major theorem provers, like
Coq,
Agda,
Abella,
Isabelle/HOL~\cite{BlanchetteM0T17},
and term-rewriting systems~\cite{EndrullisHHP015}.
The methods we described here will be applicable in many of these. For example,
we supply Coq implementation of all our running examples on the implememtation page.\cref{fn:impl}


\begin{thebibliography}{}

\bibitem[\protect\citeauthoryear{Barendregt, Dekkers, and Statman}{Barendregt
  et~al\mbox{.}}{2013}]{Barendregt:LambdaCalcTypes}
{\sc Barendregt, H.}, {\sc Dekkers, W.}, {\sc and} {\sc Statman, R.} 2013.
\newblock {\em Lambda {{Calculus}} with {{Types}}}.
\newblock {Cambridge University Press}, {Cambridge ; New York}.

\bibitem[\protect\citeauthoryear{Basold, Komendantskaya, and Li}{Basold
  et~al\mbox{.}}{2019}]{BasoldKL19-2}
{\sc Basold, H.}, {\sc Komendantskaya, E.}, {\sc and} {\sc Li, Y.} 2019.
\newblock Coinduction in uniform: Foundations for corecursive proof search with
  horn clauses.
\newblock In {\em {ESOP} 2019}. 783--813.

\bibitem[\protect\citeauthoryear{Blanchette et~al\mbox{.}}{Blanchette
  et~al\mbox{.}}{2017}]{BlanchetteM0T17}
{\sc Blanchette, J.} {\sc et~al\mbox{.}} 2017.
\newblock Foundational nonuniform (co)datatypes for higher-order logic.
\newblock In {\em {LICS}'17}. {IEEE} Computer Society, 1--12.

\bibitem[\protect\citeauthoryear{Boyer and Moore}{Boyer and
  Moore}{1979}]{Boyer79:ComputationalLogic}
{\sc Boyer, R.~S.} {\sc and} {\sc Moore, J.~S.} 1979.
\newblock {\em A Computational Logic}.
\newblock {{ACM}} Monograph Series. {Academic Press}.

\bibitem[\protect\citeauthoryear{Brotherston and Simpson}{Brotherston and
  Simpson}{2011}]{BrotherstonS11}
{\sc Brotherston, J.} {\sc and} {\sc Simpson, A.} 2011.
\newblock Sequent calculi for induction and infinite descent.
\newblock {\em JLC\/}~{\em 21,\/}~6, 1177--1216.

\bibitem[\protect\citeauthoryear{Courcelle}{Courcelle}{1983}]{Courcelle83}
{\sc Courcelle, B.} 1983.
\newblock Fundamental properties of infinite trees.
\newblock {\em {TCS}\/}~{\em 25}, 95--169.

\bibitem[\protect\citeauthoryear{Dagnino, Ancona, and E.Zucca}{Dagnino
  et~al\mbox{.}}{2020}]{DAZ20}
{\sc Dagnino, F.}, {\sc Ancona, D.}, {\sc and} {\sc E.Zucca}. 2020.
\newblock Flexible coinductive logic programming.
\newblock {\em TPLP\/}.

\bibitem[\protect\citeauthoryear{Endrullis, Hansen, Hendriks, Polonsky, and
  Silva}{Endrullis et~al\mbox{.}}{2015}]{EndrullisHHP015}
{\sc Endrullis, J.}, {\sc Hansen, H.~H.}, {\sc Hendriks, D.}, {\sc Polonsky,
  A.}, {\sc and} {\sc Silva, A.} 2015.
\newblock A coinductive framework for infinitary rewriting and equational
  reasoning.
\newblock In {\em {RTA}'15}. 143--159.

\bibitem[\protect\citeauthoryear{Fu and Komendantskaya}{Fu and
  Komendantskaya}{2016}]{FK16}
{\sc Fu, P.} {\sc and} {\sc Komendantskaya, E.} 2016.
\newblock Operational semantics of resolution and productivity in {H}orn clause
  logic.
\newblock {\em Formal Aspects of Computing\/}.

\bibitem[\protect\citeauthoryear{Fu, Komendantskaya, Schrijvers, and Pond}{Fu
  et~al\mbox{.}}{2016}]{FKS15}
{\sc Fu, P.}, {\sc Komendantskaya, E.}, {\sc Schrijvers, T.}, {\sc and} {\sc
  Pond, A.} 2016.
\newblock Proof relevant corecursive resolution.
\newblock In {\em FLOPS'16}. Springer, 126--143.

\bibitem[\protect\citeauthoryear{Gentzen}{Gentzen}{1969}]{Gentzen69}
{\sc Gentzen, G.} 1969.
\newblock Investigations into logical deduction.
\newblock In {\em The Collected Papers of Gerhard Gentzen}, {M.~Szabo}, Ed.
  Studies in Logic and the Foundations of Mathematics, vol.~55. Elsevier, 68 --
  131.

\bibitem[\protect\citeauthoryear{Gupta, Bansal, Min, Simon, and Mallya}{Gupta
  et~al\mbox{.}}{2007}]{GuptaBMSM07}
{\sc Gupta, G.}, {\sc Bansal, A.}, {\sc Min, R.}, {\sc Simon, L.}, {\sc and}
  {\sc Mallya, A.} 2007.
\newblock Coinductive logic programming and its applications.
\newblock In {\em Logic Programming}, {V.~Dahl} {and} {I.~Niemel{\"a}}, Eds.
  Springer Berlin Heidelberg, Berlin, Heidelberg, 27--44.

\bibitem[\protect\citeauthoryear{Kimura, Nakazawa, Terauchi, and Unno}{Kimura
  et~al\mbox{.}}{2020}]{KNTU20}
{\sc Kimura, D.}, {\sc Nakazawa, K.}, {\sc Terauchi, T.}, {\sc and} {\sc Unno,
  H.} 2020.
\newblock Failure of cut-elimination in cyclic proofs of separation logic.
\newblock {\em Computer Software\/}~{\em 37}, 39--52.

\bibitem[\protect\citeauthoryear{Komendantskaya and Li}{Komendantskaya and
  Li}{2017}]{KL17}
{\sc Komendantskaya, E.} {\sc and} {\sc Li, Y.} 2017.
\newblock Productive corecursion in logic programming.
\newblock {\em J. {TPLP} (ICLP'17 post-proc.)\/}~{\em 17,\/}~5-6, 906--923.

\bibitem[\protect\citeauthoryear{Miller and Nadathur}{Miller and
  Nadathur}{2012}]{MN12}
{\sc Miller, D.} {\sc and} {\sc Nadathur, G.} 2012.
\newblock {\em Programming with Higher-order logic}.
\newblock Cambridge University Press.

\bibitem[\protect\citeauthoryear{Miller, Nadathur, Pfenning, and
  Scedrov}{Miller et~al\mbox{.}}{1991}]{MNPS91}
{\sc Miller, D.}, {\sc Nadathur, G.}, {\sc Pfenning, F.}, {\sc and} {\sc
  Scedrov, A.} 1991.
\newblock {\em Uniform Proofs as a Foundation for Logic Programming}. Annals of
  Pure and Applied Logic, vol.~51.
\newblock Elsevier, 125--157.

\bibitem[\protect\citeauthoryear{Plotkin}{Plotkin}{1970}]{plotkin1970note}
{\sc Plotkin, G.~D.} 1970.
\newblock A note on inductive generalization.
\newblock {\em Machine intelligence\/}.

\bibitem[\protect\citeauthoryear{Saotome, Nakazawa, and Kimura}{Saotome
  et~al\mbox{.}}{2020}]{SNK20}
{\sc Saotome, K.}, {\sc Nakazawa, K.}, {\sc and} {\sc Kimura, D.} 2020.
\newblock Restriction on cut in cyclic proof system for symbolic heaps.
\newblock In {\em FLOPS'20}.

\bibitem[\protect\citeauthoryear{Schubert and Urzyczyn}{Schubert and
  Urzyczyn}{2018}]{SchubertU18}
{\sc Schubert, A.} {\sc and} {\sc Urzyczyn, P.} 2018.
\newblock First-order answer set programming as constructive proof search.
\newblock {\em Theory Pract. Log. Program.\/}~{\em 18,\/}~3-4, 673--690.

\bibitem[\protect\citeauthoryear{Simon, Mallya, Bansal, and Gupta}{Simon
  et~al\mbox{.}}{2006}]{SimonEtAl06}
{\sc Simon, L.}, {\sc Mallya, A.}, {\sc Bansal, A.}, {\sc and} {\sc Gupta, G.}
  2006.
\newblock Coinductive logic programming.
\newblock In {\em ICLP}. 330--345.

\bibitem[\protect\citeauthoryear{Sorensen and Urzyczyn}{Sorensen and
  Urzyczyn}{2006}]{SU06}
{\sc Sorensen, M.~H.} {\sc and} {\sc Urzyczyn, P.} 2006.
\newblock {\em Lectures on the Curry-Howard Isomorphism}.
\newblock Studies in Logic. Elsevier.

\bibitem[\protect\citeauthoryear{Sulzmann, Duck, Jones, and Stuckey}{Sulzmann
  et~al\mbox{.}}{2007}]{SulzmannDJS07}
{\sc Sulzmann, M.}, {\sc Duck, G.~J.}, {\sc Jones, S. L.~P.}, {\sc and} {\sc
  Stuckey, P.~J.} 2007.
\newblock Understanding functional dependencies via constraint handling rules.
\newblock {\em J. Funct. Program.\/}~{\em 17,\/}~1, 83--129.

\bibitem[\protect\citeauthoryear{Troelstra and Schwichtenberg}{Troelstra and
  Schwichtenberg}{2000}]{Troelstra00:BasicProofTheory}
{\sc Troelstra, A.~S.} {\sc and} {\sc Schwichtenberg, H.} 2000.
\newblock {\em Basic {{Proof Theory}}\/}, 2nd ed.
\newblock Cambridge {{Tracts}} in {{Theoretical Computer Science}}. {Cambridge
  University Press}, {Cambridge}.

\end{thebibliography}

\end{document}